\documentclass[aps,prb,reprint,superscriptaddress,flushbottom,
]{revtex4-1}

\usepackage{amsmath}
\usepackage{amssymb}
\usepackage{amsthm}

\usepackage{setspace}
\long\def\symbolfootnote[#1]#2{\begingroup\def\thefootnote{\fnsymbol{footnote}}\footnote[#1]{#2}\endgroup}
\usepackage{amscd}

\usepackage{color}

\usepackage{float}

\newcommand{\bra}[1]{\langle {#1} \vert}
\newcommand{\ket}[1]{\vert {#1} \rangle}

\newcommand{\del}[0]{\partial}

\usepackage{epsfig}

\usepackage{hyperref}
\usepackage{cleveref} 

\begin{document}
\title{Spatial distribution of superfluidity and superfluid distillation of Bose liquids}
%
\author{T.J. Volkoff}
\email{volkoff@konkuk.ac.kr}
\affiliation{Department of Physics, Konkuk University, Seoul 05029, Korea}
\author{Yongkyung Kwon}
\affiliation{Department of Physics, Konkuk University, Seoul 05029, Korea}

\begin{abstract}
Under the assumption of two fluid kinematics of a nonrelativistic Bose liquid in the presence of a local velocity field $v(x)$, local Galilei transformations are used to derive formulas for the spatial distribution of superfluidity. The local formulation is shown to subsume several descriptions of superfluidity, from Landau's free quasiparticle picture of the normal fluid to the fully microscopic winding number formula for superfluid density. We derive the superfluid distribution of generic pure states of 1-d bosonic systems by using the continuum analog of matrix product states. With a view toward spatially structured superfluid-based quantum devices, we consider the limits to local distillation of superfluidity within the framework of localized resource theories of quantum coherence.
\end{abstract}
\maketitle

\section{Introduction}
The dissipationless flow of He II and its connection to the structure of the bosonic many-body quantum state remain central problems in the study of strongly correlated bosonic systems. Among the celebrated achievements for zero temperature systems are F. London's proposal to describe the He II wavefunction via a local phase field, \cite{londonorig} Feynman's utilization of many-body quantum coherence to incorporate density fluctuations and vortices, \cite{feynmanorig1,feynmanwavefunct,feynmanvort} and Feynman and Cohen's inclusion of particle pair correlations leading to agreement with Landau's proposed excitation spectrum.\cite{feyncoh} For equilibrium He II at finite temperature, \textit{ad hoc} quantification of the local kinematic superfluidity based on the winding number or projected area of imaginary time polymers has led to a microscopic understanding of superfluidity at the nanoscale.\cite{kwonwhaley,volkoffkwonwhaley,delmaestro,royreview,draegerceperley,paesanikwonwhaley} However, despite these major theoretical advances, the local distribution of superfluidity in general classes of bosonic variational wavefunctions has not been analyzed and, furthermore, the estimators of local superfluid density based on properties of imaginary time polymers are not applicable to generic bosonic states.\cite{rousseau} Therefore, the development and interpretation of a widely applicable, quantum mechanical framework describing local superfluidity in systems that exhibit two-fluid behavior is desirable. Recent experiments demonstrating optomechanical control of excitations in He II film,\cite{optomechcontrol} and He II-based interferometry,\cite{PhysRevLett.106.255301} which indicate the possibility of local control of superfluid dynamics on a range of length scales, highlight the necessity of a local theory of superfluidity for superfluid-based devices.

In the present work, we derive a general framework for calculating the local normal fluid distribution for bosonic systems that empirically exhibit local two fluid behavior. Because the framework is based on local Galilei transformations of a generic bosonic system, it can be used to extend Landau's quasiparticle picture of the normal fluid or to analyze local superfluidity in the microscopic theory. We do not aim to derive the two fluid picture from quantum mechanical principles, which is the subject of various monographs,\cite{griffinbook,robertsberloff,bogo2} but rather determine how the superfluid kinematics (e.g., the distribution of superfluidity) depends on a given spatially varying velocity field in the system and on the structure of the bosonic quantum state. As examples, we generalize Landau's calculation of the contributions of free quasiparticle gases to the normal fluid, determine the local superfluid distribution of a $k=0$ Bose-Einstein condensate in the presence of a generic smooth velocity field, and derive a general formula based on continuum matrix product states for the local superfluid density in one spatial dimension.
Finally, we formulate a resource theory of quantum coherence with the aim of deriving fundamental limits for distillation of a perfect superfluid state in a spatial subregion of a bosonic system under a physically relevant class of quantum operations. Our results greatly extend the possibilities for numerical simulation of superfluid systems in the presence of flow, and can be applied to the design of gratings and external potentials in bosonic matter wave interferometers \cite{PhysRevLett.101.085302,PhysRevLett.110.093602,PhysRevLett.105.243003} and local heat flux control protocols for interacting bosonic liquids and gases.\cite{PhysRevLett.98.195302,PhysRevA.86.033619}

To derive a formula for the local normal fluid distribution, we first generalize Landau's quasiparticle picture of the normal fluid. In a system of $N$ atomic bosons, we consider a subset $\mathcal{S}$ of particles \footnote{These particles may be atoms themselves or bosonic degrees of freedom arising from an effective description of the atomic system.} that follow integral curves of a velocity field $v(x)$, which defines a steady flow on a compact connected subset $\Omega \subset \mathbb{R}^{d}$ with volume $\vert \Omega \vert >0$. In Landau's treatment, this subset is idealized as a gas of noninteracting bosonic quasiparticles, but we do not insist on this interpretation. According to the local two fluid model (see below), the local normal density with respect to $v(x)$ is the part of the fluid that is moving with respect to a local reference frame that has velocity $-v(x)$ with respect to the subset $\mathcal{S}$. In particular, the subset $\mathcal{S}$ contributes to the normal density, which may have contributions from other degrees of freedom, e.g., collective modes, pinned vortices, etc. To derive the unitary operator that represents the local Galilei transformation (LGT) mapping between these local reference frames, we first use the bosonic canonical commutation relation $[\psi(x),\psi(x')^{\dagger}]=\delta(x-x')$ \footnote{In this work, we omit the projection onto the symmetrized Fock space; all creation (annihilation) operations map the symmetrized Fock space of $N$ particles to the symmetrized Fock space of $N+1$ ($N-1$) particles.} to show that for any differentiable vector field $f:\mathbb{R}^{d}\rightarrow \mathbb{R}^{d}$ and for $U[f(x)]:= \exp (i\int d^{d}x\, f(x)^{T}x\, \psi(x)^{\dagger}\psi(x))$ a unitary operator-valued functional, 
\begin{eqnarray}
U[f(x)]\psi(x)U[f(x)]^{\dagger}&=&\psi(x)e^{-if(x)^{T}x}\nonumber \\
U[f(x)]\psi(x)^{\dagger}U[f(x)]^{\dagger}&=&\psi(x)^{\dagger}e^{if(x)^{T}x}.
\label{eqn:unitaryaction}
\end{eqnarray}
Taking $P:=\int d^{d}x\,g(x)$ to be the total momentum, where $g(x):=(1/2i)\psi(x)^{\dagger}(\nabla - \overset{\leftarrow}{\nabla})\psi(x)$ is the momentum density (we take $\hbar=k_{B}=1$ throughout), it then follows that (see Appendix \ref{sec:app1})
\begin{equation}
U[f(x)]P U[f(x)]^{\dagger} = P-\int d^{d}x\, (\nabla (f(x)^{T}x)) \psi(x)^{\dagger}\psi(x).
\label{eqn:ptransf}
\end{equation}
The connection to the usual Galilei transformation \cite{takahashi,penco,hohenberg,kambe} is made by choosing $f(x)=mv$ to be a constant momentum in Eq.(\ref{eqn:ptransf}), where $m$ is the atomic mass. However, in the present work we do not make the assumption of homogeneity, and instead allow $f(x)=mv(x)$, where $v(x)$ is a smooth vector field on $\mathbb{R}^{d}$. We note that the function $\sum_{\ell=1}^{N}mv(R^{(\ell)})^{T}R^{(\ell)}$, where $R^{(\ell)}$ is the position of the $\ell$-th atom, was proposed by Feynman to give the phase of a superfluid state of $N$ atoms with slowly-varying velocity (see Eq.(16) of Ref.\onlinecite{feynmanvort}).

An outline of the paper is as follows: Section \ref{sec:localsf} reviews the traditional approach to calculation of a global value of the normal fluid density and introduces the local framework for calculation of the normal fluid distribution based on the LGT in Eq.(\ref{eqn:unitaryaction}). In Subsection \ref{sec:qpandbec} we apply the local framework to the setting of quasiparticle gases and Bose-Einstein condensed states. In Subsection \ref{sec:wind}, a local generalization of the winding number estimator of superfluid density is derived, which provides a link to numerical calculations of strongly interacting bosonic systems. Section III contains a calculation of the normal fluid distribution for all bosonic states in one spatial dimension, which is particularly useful when the state has an efficient approximation by matrix product states. In Section IV, a localized resource theory of quantum coherence is defined which allows superfluidity to be considered as a local quantum resource, giving an information-based context for superfluid distillation protocols based on, e.g., thermo-mechanical effects.

\section{\label{sec:localsf}Local superfluidity} Given a system of $N$ bosons of mass $m$ at thermal equilibrium at temperature $\beta^{-1}$ and density $\rho={mN\over \vert \Omega \vert}$, where  $\Omega$ is defined as before, a standard derivation of the global normal fluid tensor consists of the following recipe: \cite{landaustatphys2,khalatbook,schmittbook,rousseau,pines} 1) the two fluid assumption $(\rho_{s})_{i,j}=\rho -(\rho_{n})_{i,j}$, 2) calculation of the expectation of $P_{i}$ in the Galilei boosted Gibbs state $\sigma(\beta)_{v_{j}}:=U[mv_{j}]e^{-\beta H}U[mv_{j}]^{\dagger}/\text{tr}e^{-\beta H}$, 3) use of the two fluid assumption and the correspondence principle to equate the observed momentum to the expectation value of the momentum
\begin{equation}
(\rho_{n})_{i,j}v_{j}\vert \Omega \vert = \text{tr}P_{i}\sigma(\beta)_{v_{j}}.
\label{eqn:normalusual}
\end{equation}
In the limit of zero relative momentum between superfluid and normal fluid fractions, one has $(\rho_{n})_{i,j}\vert_{v_{j}=0}=\lim_{v_{j}\rightarrow 0} (\vert \Omega \vert v_{j})^{-1}\text{tr}P_{i}\sigma(\beta)_{v_{j}}$.

A local, equilibrium version of the normal fluid tensor in the presence of a constant relative velocity between the normal and superfluid components can be formulated in terms of the momentum density $g(x)$.\cite{baym} We presently generalize this approach to include a spatially varying relative velocity field $v(x)$ and an arbitrary bosonic quantum state associated with positive, trace class operator $\sigma$. The analogous framework is as follows: 1') the local two-fluid assumption is given by $\rho_{s}(x)_{i,j}:=\text{tr}\psi(x)^{\dagger}\psi(x)\sigma -\rho_{n}(x)_{i,j}$ for all $x\in \Omega$, 2') calculation of the expectation of $g(x)_{i}$ in the locally Galilei transformed state $\sigma_{v(x)_{j}}:= U[mv(x)_{j}]\sigma U[mv(x)_{j}]^{\dagger}$, 3') use of the local two fluid assumption and the local correspondence principle to equate the observed momentum density to the expectation value of the momentum density 
\begin{equation}
(\rho_{n}(x))_{i,j}v(x)_{j} = \text{tr}g(x)_{i}\sigma_{v(x)_{j}}.
\label{eqn:nolimit}
\end{equation}

In particular, prescription 3') leads to the following formula for the local normal fluid tensor in the limit $v(x)\rightarrow 0$:
\begin{equation}
\rho_{n}(x)_{i,j}\big\vert_{v(x)_{j}=0}=\lim_{v(x)_{j}\rightarrow 0}v(x)_{j}^{-1}\text{tr}g(x)_{i}\sigma_{v(x)_{j}}.
\label{eqn:localnormal}
\end{equation}
We emphasize  that the localized two fluid assumptions 1')-3') are kinematic in nature, i.e., they are well-defined without any assumption of the quantum dynamics of $\sigma$ and, therefore, without the assumption of the existence of a Landau critical velocity in the fluid.\cite{khalatbook}  Calculations of time-dependent superfluid response, which have been carried out in certain cases for both noiseless and dissipative quantum evolution,\cite{arovas,keeling} can also be suitably localized.

\subsection{Quasiparticle gases and Bose-Einstein condensates\label{sec:qpandbec}}
As an application of Eq.(\ref{eqn:nolimit}), we outline the LGT generalization of Landau's formula for the contribution to $\rho_{n}$ of a non-interacting quasiparticle gas in equilibrium,\cite{landaustatphys2,landauorig} leaving detailed examples to a future work. Under an LGT in the $j$-th direction, the system Hamiltonian $H$ is transformed to $H-H' + \mathcal{O}(v(x)^{2})$, where $H':=\int d^{d}x \, \left( g(x)^{T}\nabla (v(x)_{j}x_{j}) - {m\over 2}\Vert \nabla (v(x)_{j}x_{j}) \Vert^{2}\psi(x)^{\dagger}\psi(x) \right)$ (see Eq.(\ref{eqn:lgtham} of Appendix \ref{sec:windingproof}). Taking the spectrum of the quasiparticle gas to be $\epsilon(k)$, one finds that application of the LGT to the system Gibbs state produces a bosonic Gaussian state of the quasiparticles proportional to $e^{-\beta(\sum_{k}\epsilon (k)a_{k}^{\dagger}a_{k} - H')}$, which reduces to $e^{-\beta \sum_{k}(\epsilon(k)-k^{T}v)a_{k}^{\dagger}a_{k}}$ for $v(x)=v$. The right side of Eq.(\ref{eqn:nolimit}) can then be calculated by standard methods, e.g., Wick's theorem. 

The LGT in Eq.(\ref{eqn:unitaryaction}) is defined for any bosonic quantum state in the domain of $\int d^{d}x \, (f(x)^{T}x)\psi(x)^{\dagger}\psi(x)$. Therefore, formulas (\ref{eqn:nolimit}), (\ref{eqn:localnormal}) do not require that a large fraction of particles occupy a single-particle state. Of course, as a special case, the present analysis can be used to determine the conditions for which a local two-fluid description is valid for a state that exhibits homogeneous Bose-Einstein condensation (BEC) of excitations. For example, whereas several previous calculations of kinematic superfluid density making use of global Galilei transformations assign $\rho_{n}\vert_{v_{j}=0}=0$ to the zero momentum BEC state $\ket{\psi_{k=0}}:={1\over \sqrt{N!}}a_{k=0}^{\dagger N}\ket{\text{VAC}}$,\cite{baym,liebbook} we obtain the following more descriptive result, the derivation of which is given in Appendix \ref{sec:becvproof}: 
\begin{equation}
\rho_{n}(x)_{i,j}\big\vert_{v(x)_{j}=0} = {Nm\over \vert \Omega \vert}\delta_{i,j} + {Nm\over \vert \Omega \vert}\lim_{v(x)_{j}\rightarrow 0} {x_{j}\del_{i}v(x)_{j} \over v(x)_{j}}.
\label{eqn:becresult}
\end{equation}
In the case of $v(x)=\omega(-x_{2},x_{1})$ on $\mathbb{R}^{2}$, i.e., a rotating film with angular frequency $\omega$, Eq.(\ref{eqn:becresult}) implies a complete longitudinal response $\rho_{n}(x)_{j,j}=mN/ \vert \Omega \vert$. By contrast, the vanishing transverse response implies complete superfluidity $\rho_{s}(x)_{i,j}=mN/\vert \Omega \vert$, in agreement with the analysis of the rotating bucket experiment by linear reponse methods.\cite{baym}  In the more general case of an inhomogeneous BEC, Eq.(\ref{eqn:becresult}) does not hold. In fact, in Section \ref{sec:localdistill}, we derive an inhomogeneous BEC that satisfies $\rho_{n}(x)_{i,j}=0$ exactly for a given $v(x)$, that is, we derive the form of an inhomogeneous BEC that is an isotropic superfluid.

\subsection{Generalized winding number estimator\label{sec:wind}}
In \textit{ab initio} numerical calculations of He II in equilibrium, no assumption is made regarding the marginal quantum state of quasiparticles arising from the microscopic theory. Instead, the superfluid density is calculated from a Gibbs state of the entire system from the second moment of a classical random variable describing the sum of the winding numbers of the imaginary time paths that are in one-to-one correspondence with the He atoms.\cite{pollockceperley,ceperley} In order to generalize this estimation method to allow for calculation of the spatial distribution of superfluidity, we take the field operator $\psi(x)$ ($\psi(x)^{\dagger}$) in Sec.\ref{sec:localsf} to represent annihilation (creation) of a $^{4}$He atom, i.e., the constituent boson, at the position $x$ in the spatial domain. In this microscopic description, the LGT given by Eq.(\ref{eqn:unitaryaction}) is applied to the field of the constituent bosons. Denoting $\ket{R}:= \ket{x^{(1)}}_{1}\otimes \cdots \otimes \ket{x^{(N)}}_{N}$ for a quantum state of distinguishable particles at positions $\lbrace x^{(\ell)} \rbrace_{\ell=1,\ldots ,N} \subset \mathbb{R}^{d}$ \footnote{This symbol is to be considered as a limit of normalizable quantum states.}, and letting $\ket{sR}:= \ket{x^{(s(1))}}_{1}\otimes \cdots \otimes \ket{x^{(s(N))}}_{N}$ be its image under the action of an element $s$ of the symmetric group $\mathfrak{S}_{N}$, one finds that the locally Galilei transformed Gibbs state of $N$ bosons is given in the particle position basis by the kernel
\begin{scriptsize}
\begin{equation}
{1\over N!}\sum_{s\in \mathfrak{S}_{N}}e^{im\sum_{\ell=1}^{N}(v\left( x^{(\ell)}\right)^{T} x^{(\ell)}-v\left( x'^{(s(\ell))} \right)^{T}  x'^{(s(\ell))})  } \langle R \vert e^{-\beta H} \vert sR' \rangle
\label{eqn:lgtexpect}
\end{equation}
\end{scriptsize}
where $H$ is the system Hamiltonian (e.g., defined by the sum of kinetic energy and a suitable two-body interaction).  In Eq.(\ref{eqn:lgtexpect}), the presence of only a single sum over permutations is valid as long as $H$ commutes with the projector to the symmetrized Fock space.  Under a LGT, the change in the Helmholtz free energy satisfies $e^{-\beta \Delta F}=Z^{-1}{1\over N!}\sum_{s\in \mathfrak{S}_{N}}\mathbb{E}_{s} e^{ im\sum_{\ell=1}^{N}(v\left( x^{(\ell)}\right)^{T} x^{(\ell)}-v\left( x^{(s(\ell))}\right)^{T} x^{(s(\ell))})}$, where $\mathbb{E}_{s}$ is the non-normalized expectation with respect to density $\langle R \vert e^{-\beta H}  \vert sR \rangle$ on $\mathbb{R}^{Nd}$ and $Z=\text{tr}e^{-\beta H}={1\over N!}\sum_{s\in \mathfrak{S}_{N}}\mathbb{E}_{s}(1)$. Taking an LGT in the $j$-th direction and assuming that $v(x)_{j}$ is small and that the function $v(x)_{j}x_{j}$ varies slowly throughout the system, one can write $\beta \Delta F \approx {m^{2}\over 2Z} {1\over N!}\sum_{s\in \mathfrak{S}_{N}}\mathbb{E}_{s}W_{s}(\lbrace x^{(\ell)} \rbrace)_{j}^{2}$, where $W_{s}(\lbrace x^{(\ell)} \rbrace)_{j}:= \sum_{\ell=1}^{N}\nabla (v(x)_{j}x_{j})\big\vert_{x=x^{(\ell)}}^{T} (x^{(\ell)}- x^{(s(\ell))})  $. On the other hand, the constitutive relation between $\langle g(x)_{i} \rangle_{\rho(\beta)_{v(x)_{j}}}$ and $\Delta F$ is calculated from the Hamiltonian by making use of the commutation relation Eq.(\ref{eqn:unitaryaction}) (see Appendix \ref{sec:windingproof} for details). Therefore, by Eq.(\ref{eqn:nolimit}), the following formula holds:
\begin{small}
\begin{eqnarray}
\rho_{n}(x)_{i,j}&\approx  &m{\del_{i}(v(x)_{j}x_{j})\over v(x)_{j}}\langle \psi(x)^{\dagger}\psi(x) \rangle_{\rho(\beta)_{v(x)_{j}}}  \nonumber \\
&-& {m^{2}\over 2\beta \hbar^{2}N!Z v(x)_{j}}{\delta \left( \sum_{s\in\mathfrak{S}_{N}}\mathbb{E}_{s}W_{s}(\lbrace x^{(\ell)} \rbrace)_{j}^{2} \right) \over \delta \del_{i}(v(x)_{j}x_{j})} 
\label{eqn:genwind}
\end{eqnarray}
\end{small}
where $\hbar$ has been reintroduced for ease of comparison to the original winding number estimator of superfluid density,\cite{pollockceperley} which is obtained from Eq.(\ref{eqn:genwind}) for constant $v(x)=v$. The expectation value in Eq.(\ref{eqn:genwind}) can be calculated using, e.g., path integral Monte Carlo techniques.

\section{Superfluidity of continuum matrix product states} 

In Sec. \ref{sec:wind}, we obtained a general formula for the distribution of superfluidity for bosons in equilibrium in any spatial dimension. For general quantum states, it is of interest to determine a simple expression for the right hand side of Eq.(\ref{eqn:nolimit}).  Recent path integral Monte Carlo calculations of R\'{e}nyi entanglement entropy have shown that the superfluid behavior of liquid He II arises from quantum states that satisfy the area law for entanglement entropy.\cite{herdmanarea}  Due to the fact that a pure state of a finite lattice quantum system obeys the area law only if it is approximable by a matrix product state of finite bond dimension,\cite{hastings,verstraetecirac2007} one expects that quantum states of continuum quantum systems that obey the area law for R\'{e}nyi entanglement entropy can be expressed as continuum matrix product states (cMPS) in the appropriate spatial dimension. This expectation has been established rigorously in one spatial dimension.\cite{huang} By way of example, it is known that the ground state of the Lieb-Liniger model, describing one-dimensional bosons with a repulsive pair interaction given by $V(x-x')=g\delta(x-x')$, $g>0$, is well-approximated by cMPS.\cite{verstraetecirac} The Lieb-Liniger system has been experimentally realized in ultracold $^{87}$Rb confined in parallel optical tubes.\cite{weiss} In general, by calculating the right hand side of Eq.(\ref{eqn:nolimit}) for generic cMPS, we derive a formula (see Eq.(\ref{eqn:onednormal}) below) that can be directly compared to the microscopic result Eq.(\ref{eqn:genwind}) in the limit of zero temperture, which can be calculated in turn by, e.g., path-integral ground state methods.\cite{magro}

A cMPS $\ket{\psi(R,Q)}$ on the circle $[-{L\over 2},{L\over 2}]$ (with periodic boundary condition assumed) is defined by two auxiliary $D\times D$ matrix-valued functions $Q(x)$ and $R(x)$ via $\ket{\psi(R,Q)}:=\text{Tr}\mathcal{P}\exp \left( \int_{-{L\over 2}}^{{L\over 2}}dx\, Q(x)\otimes \mathbb{I}_{D} + R(x)\otimes \psi(x)^{\dagger} \right)\ket{\text{VAC}}$, where $\text{Tr}$ is the trace in the $D^{2}$-dimensional auxiliary matrix space and $\mathcal{P}$ is the path ordering operator. In order to calculate $\rho_{n}(x)$ for a cMPS $\ket{\psi [Q,R]}$, it suffices to utilize the formula for the one-body density $\rho^{(1)}_{Q,R}(x,y):=\langle \psi(x)^{\dagger}\psi(y) \rangle_{\ket{\psi [Q,R] } }$ \cite{haegemancalc} and note that $\langle g(x) \rangle_{U[mv(x)]\ket{\psi [Q,R]}} = {1\over 2i}\lim_{x\rightarrow y}\left( {d\over dy}-{d\over dx} \right) \rho^{(1)}_{Q,R}(x,y)$. Working with the gauge $Q(x)=0$ and assuming that the ground state $\ket{\psi(Q,R)}$ of the quasiparticles has no momentum at any point, the result is given by
 \begin{equation}
\langle g(x) \rangle_{U[mv(x)]\ket{\psi [Q,R]}} = \text{Tr}T\left( -{L\over 2},{L\over 2}\right) A(x)
\label{eqn:cmpsmomentumdens}
\end{equation} where $T(y,x):=\mathcal{P}\exp\left( \int_{y}^{{x}}dz\, R(z)\otimes \overline{R(z)} \right)$ and $A(x):= m\left( x{dv \over dx} +v(x) \right)R(x)\otimes \overline{R(x)}$. 
Referring to Eq.(\ref{eqn:localnormal}) and noting that the local density $\rho(x):=\langle \psi(x)^{\dagger}\psi(x) \rangle$ in cMPS is given by $\rho(x)=\text{Tr}T\left( -{L\over 2},{L\over 2}\right) R(x)\otimes \overline{R(x)}$, we find the following general formula for the velocity-dependent normal fluid density in 1-d:
\begin{equation}
\rho_{n}(x)=m\rho(x) + {mx\over v(x)}{dv \over dx} \rho(x).
\label{eqn:onednormal}
\end{equation}
The similarity of Eq.(\ref{eqn:becresult}) to  Eq.(\ref{eqn:onednormal}) belies the fully quantum nature of the latter result. Because the set of MPS is dense in the Fock space, Eq.(\ref{eqn:onednormal}) is an exact equation which can be utilized in calculations of the local superfluid density even in non-translation invariant, quasi-one-dimensional systems of He II, which have been shown to exhibit Luttinger liquid behavior.\cite{markicglyde,affleckmaestro}

Although systems of definite particle number are not described by $D=1$ cMPS, the right hand side of Eq.(\ref{eqn:onednormal}) is still useful as a generating function for $\rho_{n}(x)$. As an example, note that given a sub-Hilbert space $\mathcal{K}$ of $L^{2}([-L/2,L/2])$, any state of the form \footnote{All pure quantum states obtained from Bose symmetrization of non-entangled pure states, e.g., Bose-Einstein condensates or fragmented Bose-Einstein condensates, have this form.} \begin{equation}\prod_{k=1}^{M}{a_{\ket{\psi_{k}}}^{\dagger n_{k}}\over \sqrt{n_{k}!}}\ket{\text{VAC}}\label{eqn:frag}\end{equation} where $\sum_{k=1}^{M}n_{k}=N$ and where the canonical boson operators $a_{\ket{\psi}}$, $a_{\ket{\varphi}}^{\dagger}$ satisfy $[a_{\ket{\psi}},a_{\ket{\varphi}}^{\dagger}]=\langle \psi \vert \varphi \rangle $ for any $\ket{\psi}$, $\ket{\varphi}\in\mathcal{K}$,\cite{bratteli2} can be generated from a cMPS of bond dimension $D=1$ with $R(x)=\sum_{j=1}^{M}\xi_{j}\psi_{j}(x)$, where $\psi_{j}(x):=\langle x \vert \psi_{j} \rangle$ (see Appendix \ref{sec:cmpsapp}). Using the gauge freedom of cMPS to take $Q(x)=0$, the normal fluid distribution is given by 
\begin{footnotesize}
\begin{eqnarray}
\rho_{n}(x)&=& v(x)^{-1}\left(\prod_{j=1}^{M} \del_{\overline{\xi}_{j}}^{n_{j}}\del_{\xi_{j}}^{n_{j}} \right) \langle g(x) \rangle_{U[mv(x)]\ket{\psi[R]}}\big\vert_{\xi_{j},\overline{\xi}_{j}=0} \nonumber \\
&=& \left( m + {mx\over v(x)}{dv \over dx} \right) \left( \prod_{j=1}^{M}\del_{\overline{\xi}_{j}}^{n_{j}}\del_{\xi_{j}}^{n_{j}}\right) \rho(x)\big\vert_{\xi_{j},\overline{\xi}_{j}=0}.
\end{eqnarray}
\end{footnotesize}

Despite the fact that cMPS are dense in the bosonic Fock space generated by $L^{2}([-L/2,L/2])$,\cite{haegemanfieldstates} they do not provide an economical (i.e., low bond dimension) expression for general quantum phases of systems of pairwise interacting bosons. In particular, a more general class of states than Eq.(\ref{eqn:frag}) that incorporates pair correlations can be written as \begin{equation}
\prod_{\ell=1}^{N/2}U^{(\ell)}\left(  \sum_{j=1}^{\infty}\lambda^{(\ell)}_{j}a_{\ket{e_{j}^{(\ell)}}}^{\dagger 2} \right)^{r_{\ell}} U^{(\ell )\dagger} \ket{\text{VAC}}
\label{eqn:pcs}
\end{equation} where $U^{(\ell)}$ is a particle number conserving unitary generated by a quadratic Hamiltonian, $\sum_{\ell=1}^{N/2}r_{\ell}=N/2$, $r_{\ell}\in \mathbb{Z}_{\ge 0}$, and the use of a set of orthonormal bases $\lbrace \ket{e_{j}^{(\ell)}}_{j=1,2,\ldots} \rbrace_{\ell}$ for $\mathcal{K}$ is justified by the Schmidt decomposition for arbitrary two particle pure states. The states in Eq.(\ref{eqn:pcs}) are not in general approximable by cMPS of small $D$. However, they appear as ground states of a class of exactly solvable bosonic models,\cite{richardson} as probe states for quantum estimation of bosonic Hamiltonians,\cite{volkoffbosonicmetro} and as variational ans\"{a}tze for ground states of interacting Bose liquids and gases of a definite number of particles. In the case of a single nonzero $r_{\ell}$, the correlation functions and large-$N$ asymptotics of Eq.(\ref{eqn:pcs}) have been calculated.\cite{jiangcaves} Although calculation of the local superfluid density of the class of states defined by Eq.(\ref{eqn:pcs}) is beyond the scope of the present work, we note that the methods used to obtain, e.g., Eq.(\ref{eqn:becresult}), are applicable to any Gaussian state of the quasiparticle field and, therefore, to any projection of a Gaussian state to a finite particle number sector. For example, for $r_{\ell}={N\over 2}\delta_{\ell,1}$, $\ket{e^{(1)}_{j}}:=\ket{2\pi j \over L}$ a single particle momentum eigenstate (the sum over $j$ is now taken from $-\infty$ to $\infty$), and $U^{(1)}=\prod_{k>0}e^{-i{\pi \over 4}(a_{k}^{\dagger}a_{-k}+h.c.)}e^{i{\pi\over 4}(a_{k}^{\dagger}a_{k}+a_{-k}^{\dagger}a_{-k})}$ with $k\in {2\pi \over L} \mathbb{Z}$,  Eq.(\ref{eqn:pcs}) is simply the projection to fixed depletion number $N$ of the $k\neq 0$ Bogoliubov ground state $e^{-\sum_{k>0}\alpha_{k}a_{k}^{\dagger}a_{-k}^{\dagger}}\ket{\text{VAC}}$ (which is Gaussian, with $\alpha_{k={2\pi j \over L}}=: \lambda_{\pm j }^{(1)}$ determined by Bogoliubov transformation).\cite{ueda}. The normal fluid distribution of the Bogoliubov ground state can then be straightforwardly computed by applying an LGT and subsequently calculating Eq.(\ref{eqn:nolimit}) using the same method as for thermal Gaussian states.

\section{Limits on distillation of superfluidity\label{sec:localdistill}} 

Phenomena in He II such as the fountain effect and dissipationless flow through nanoporous Vycor packing provide examples of quantum dynamics that result in the conversion of partially superfluid system of $N$ atoms to a system of $N'<N$ atoms with greater superfluid fraction. As superfluid devices such as matter wave interferometers become controllable by local modification of quasiparticle states and dynamics, the ultimate limits to such distillation protocols, taking into account potentially very many system copies, becomes relevant. Furthermore, it is known from early studies that the flow of He II through capillaries of diameter $\mathcal{O}(1\text{ nm})$ constitutes an ``entropy filter'',\cite{london,pines} but a quantum information-based analysis of this phenomenon has not been carried out. However, recent axiomatic formulation of a \textit{resource theory of quantum coherence} (RTQC) allows to calculate asymptotic concentration and dilution rates between quantum states in terms of coherence measures.\cite{plenio}  Given a fixed total number of particles $N$, the form of the LGT and Eq.(\ref{eqn:nolimit}) allow to identify quantum states of $N'<N$ particles that exhibit maximal superfluid behavior in a sub-region of the system domain in a sense that we make precise below. By equating maximal superfluidity with a maximal resource in an RTQC, one can rigorously consider the limits to formation and distillation of superfluid systems. We note that modern quantum technologies based on ultracold bosonic or fermionic atoms trapped in optical lattices often utilize several thousands of individually controllable spatial subregions, so the asymptotic results of RTQC are not necessarily as remote as they may appear. Before deriving the form of the maximal local superfluid resource based on the LGT of Eq.(\ref{eqn:unitaryaction}), we first provide a concise definition of an RTQC which is specialized to the present purpose (a general definition of quantum resource theories can be found in Ref.\cite{chitambarqrt}).  

An RTQC is defined by a 4-tuple $(\mathcal{B} , \Delta , \mathcal{T}, C)$, where $\mathcal{B}=\lbrace \ket{f_{j}}\rbrace_{j\in J}$ in an orthonormal set in Hilbert space $\mathcal{H}$, $\Delta$ is the set of probability measures over the set of projections $\lbrace \ket{f_{j}}\bra{f_{j}} \rbrace_{j\in J}$ (i.e., $\Delta$ is the set of \textit{incoherent} quantum states), $\mathcal{T}$ is the set of completely positive, trace preserving maps $\Phi$ such that $\Phi(\Delta)\subset \Delta$, and $C$ is a coherence monotone that satisfies the RTQC axioms. 

To apply RTQC with the aim of deriving the rates for distillation of maximally superfluid states in a subdomain $\Omega$ of the system, we first define for a region $\Omega$ and bosonic quantum state $\sigma$ the quantity $N_{\Omega}:=\lfloor \int_{\Omega}d^{d}x\langle \psi(x)^{\dagger}\psi(x) \rangle_{\sigma}\rfloor$. For any value of $N_{\Omega}$, we define $\mathcal{B}_{\Omega}:= \lbrace \prod_{j=1}^{N_{\Omega}}\psi(x_{j})^{\dagger}\ket{\text{VAC}}: x_{j}\in \Omega \rbrace$ and use an index $I$ to label elements of $\mathcal{B}_{\Omega}$ \footnote{In practice, $\mathcal{B}_{\Omega}$ can be made to span a finite dimensional Hilbert space by using a fine spatial grid on $\Omega$.}.  One can clearly see that any $\ket{ \phi_{i}} \in \mathcal{B}_{\Omega}$, $i\in I$, has $\rho_{s}(x)_{j,j'}=0$ for all $x\in \Omega$. For any bosonic state $\sigma$ of $N$ particles, one defines $\sigma_{\Omega}:= \sum_{i,i'\in I}\langle \phi_{i} \vert \text{tr}_{N\setminus \lbrace 1,\ldots , N_{\Omega}\rbrace}\sigma \vert \phi_{i'} \rangle \ket{\phi_{i}}\bra{\phi_{i'}}$ to be the marginal state corresponding to $N_{\Omega}$ particles in $\Omega$, where $\vert \Omega \vert$ is the volume of $\Omega$. By defining the decohering map $\Delta_{\Omega}(\sigma_{\Omega}) := \sum_{i\in I}\langle \phi_{i} \vert \sigma_{\Omega} \vert \phi_{i}\rangle \ket{\phi_{i}} \bra{\phi_{i}}$, one forms the coherence measure $C_{\Omega}(\sigma):=S(\Delta(\sigma_{\Omega}))-S(\sigma_{\Omega})$, where $S$ is the von Neumann entropy of a quantum state. By defining the Hilbert spaces $\mathcal{H}_{\Omega}:=\overline{\text{span}_{\mathbb{C}} \mathcal{B}_{\Omega}}$, and associating to any relatively open set $\Omega' \subset \Omega$ the product of linear maps $P_{N_{\Omega'}} P_{\Omega',\Omega}:\mathcal{H}_{\Omega}\rightarrow \mathcal{H}_{\Omega'}$, where $P_{\Omega',\Omega}:=\int_{\Omega '}d^{d}x\, \otimes _{j=1}^{N_{\Omega }}\ket{x_{j}}_{j}\bra{x_{j}}_{j}$ is the domain projection to $\Omega '$ and $P_{N}$ is the projection to the $N$ particle sector, one obtains a sheaf of RTQCs over the system domain.

With these definitions in place, RTQC theorems can now be applied to calculate the asymptotic rate of distillation of  superfluid states in $\Omega$. In the following, we use the term \textit{strongly superfluid state} to mean a pure state such that the action of $U[mv(x)]$ on that state gives a zero eigenvector of $g(x)$. Given a velocity field $v(x)$, a rectangular subregion $\Omega$, and the state $\sigma$, Eq.(\ref{eqn:unitaryaction}), (\ref{eqn:nolimit}) imply that the state \begin{equation}\ket{\text{SF}_{\Omega}}\propto \left( {1\over \sqrt{ \vert \Omega \vert}} \int_{\Omega} d^{d}x\, e^{-imv(x)^{T}x}\psi(x)^{\dagger} \right)^{N_{\Omega}}\ket{\text{VAC}}\label{eqn:strongsf}\end{equation} is strongly superfluid in $\Omega$ (viz., because it is defined so that the action of $U[mv(x)]$ produces a $k=0$ Bose-Einstein condensate of $N_{\Omega}$ particles in $\Omega$), and it is readily verified that $C_{\Omega}(\ket{\text{SF}_{\Omega}})=N_{\Omega}\log_{2}\vert \Omega \vert$. The state $\ket{\text{SF}_{\Omega}}$ is also a maximally coherent state of the RTQC. It now follows from the basic distillation theorem of RTQC that for any $\epsilon >0$, there is an incoherent operation $T$ and an $m\in \mathbb{N}$ such that $\Vert T(\sigma_{\Omega}^{\otimes n}) - \ket{\text{SF}_{\Omega}}\bra{ \text{SF}_{\Omega}}^{\otimes nR} \Vert_{1} < \epsilon$ for all $n>m$ if and only if $R< \lfloor C_{\Omega}(\sigma) /  N_{\Omega}\log_{2}\vert \Omega \vert \rfloor$, where $\Vert \cdot \Vert_{1}$ is the trace norm.\cite{PhysRevA.92.022124,winter} The rate $R$ gives the ultimate limit for distillation of a perfect superfluid in independent copies of the domain $\Omega$ from independent copies of a marginal state of a Bose liquid, under potentially noisy quantum operations that, roughly speaking, map solid phases to solid phases.

\section{Discussion \label{sec:disc}}
Given a velocity field $v(x)$, we have utilized local Galilei transformations to develop a framework for the analysis of local kinematic superfluidity for quantum states of a wide variety of bosonic systems. If the system exhibits a basic local two fluid behavior in the sense that a local subset of particles moves along integral curves of $v(x)$ while the complementary subset remains approximately motionless, then $v(x)$ can be considered as the velocity $v_{n}(x)$ of the local normal component. Experimental methods for inducing a velocity field in interacting bosonic systems range from pressure differentials across aperture arrays separating reservoirs of liquid $^{4}$He,\cite{volkoffwhaley} to stimulated Raman adiabatic passage-based imprinting of an optical phase field on ultracold atomic Bose gases.\cite{dowlingvort}

The present results set the stage for further numerical and analytical calculations of local superfluidity. For example, the local generalization of the winding number estimator of superfluid density (which follows from Eq.(\ref{eqn:genwind})) allows to calculate the superfluid distribution via path integral Monte Carlo techniques. On the other hand, we utilized the same framework to construct the local RTQC in Section \ref{sec:localdistill}, which provides a way to consider the distillation of superfluidity in the context of modern quantum information theory. We note that our assumption in the present work of a single defining velocity field is a simplification. Realistic systems of interacting bosons may exhibit such phenomena as boundary adsorption, free surface dynamics, or phase coexistence that require a description in terms of more than one motional mode. Mathematically, one could envision generalizations of the state in Eq.(\ref{eqn:strongsf}) of the form $\ket{\psi(\lbrace \Omega_{j} \rbrace, \lbrace v_{j}(x) \rbrace)} \propto \prod_{j}\left( \int_{\Omega_{j}} d^{d}x\, e^{-imv_{j}(x)^{T}x}\psi(x)^{\dagger} \right)^{N_{\Omega_{j}}}\ket{\text{VAC}}$, such that $\sum_{j}N_{\Omega_{j}}=N$. In addition to these more complicated situations, extension of the present approach to spin-dependent flows accessible in systems of spinor bosons,\cite{PhysRevB.80.024420} and the quantum mechanical treatment of localized superfluidity for relativistic two-fluid systems,\cite{khalatnikovcarter} constitute two of several potential avenues of future research.

\acknowledgements

This work was supported by the Korea Research Fellowship Program through the National Research Foundation of Korea (NRF) funded by the Ministry of Science and ICT (2016H1D3A1908876) and by the Basic Science Research Program 
through the NRF funded by the Ministry of Education (2015R1D1A1A09056745).

\appendix
\begin{widetext}
\section{Proof of Eq.(2)\label{sec:app1}}

The proof utilizes the commutation relations $[ \psi(x),\nabla ' \psi(x')^{\dagger}]=\nabla ' \delta(x-x')$ and $[ \psi(x)^{\dagger} ,\nabla ' \psi(x')] = -\nabla ' \delta(x-x')$. In detail, by defining $A:= \int d^{d}x \, f(x) \psi(x)^{\dagger}\psi(x)$, and using $P=-i\int d^{d}x\, \psi(x')^{\dagger}\nabla ' \psi(x')$, it follows that

\begin{eqnarray}
AP-PA&=& -i \int d^{d}x\int d^{d}x' \, f(x)\left( \delta(x-x')\psi(x)^{\dagger}+\psi(x')^{\dagger}\psi(x)^{\dagger}\psi(x) \right) \nabla ' \psi(x') - PA \nonumber \\
&=& -i\int d^{d}x\, f(x)\psi(x)^{\dagger}\nabla \psi(x) -i \int d^{d}x\int d^{d}x' \,f(x)\psi(x')^{\dagger}\left( -(\nabla ' \delta(x-x'))\psi(x) + (\nabla ' \psi(x') )\psi(x)^{\dagger}\psi(x) \right) - PA \nonumber \\
&=& -i\int d^{d}x\, f(x)\psi(x)^{\dagger}\nabla \psi(x)+i \int d^{d}x\int d^{d}x' \,f(x)\psi(x')^{\dagger} (\nabla '\delta(x-x'))\psi(x) \nonumber \\
&=& i\int d^{d}x\, (\nabla f(x)) \psi(x)^{\dagger}\psi(x)
\end{eqnarray}

where the last line follows from integration by parts. Then, since $[\int d^{d}x \, f_{1}(x)\psi(x)^{\dagger}\psi(x), \int d^{d}x \, f_{2}(x)\psi(x)^{\dagger}\psi(x)]=0$ for any functions $f_{1}$, $f_{2}$, it is clear that $e^{iA}Pe^{-iA}=P- \int d^{d}x\, (\nabla f(x)) \psi(x)^{\dagger}\psi(x)$.$\square$

\section{Proof of Eq.(6)\label{sec:becvproof}}

To prove Eq.(6), we specialize to the box $\Omega = [-{L\over 2},{L\over 2}]^{\times d}$.  The action of an LGT on the creation operator $a_{k}^{\dagger}$ is given by:
\begin{equation}
U[mv(x)]a_{k}^{\dagger}U[mv(x)]^{\dagger}= {1\over L^{d/2}}\int d^{d}x \, e^{i(mv(x)^{T}x + k^{T}x)}\psi(x)^{\dagger}.
\end{equation}
It then follows that $U[mv(x)]\ket{\psi_{k=0}} = {1\over \sqrt{N!}}\left( \sum_{k}c_{k}a_{k}^{\dagger} \right)^{N}\ket{\text{VAC}}$, where
\begin{eqnarray}
c_{k}&=&{1\over L^{d}}\int d^{d}x\, e^{i(mv(x)-k)^{T}x} \nonumber \\
&=& \delta_{k,0}+ i{m\over L^{d}}\int d^{d}x\, v(x)^{T}x e^{-ik^{T}x} + \mathcal{O}(v^{2}) .
\end{eqnarray}
The expectation value of $g(x)$ in a state of the form ${1\over \sqrt{N!}}\left( \sum_{k}c_{k}a_{k}^{\dagger} \right)^{N}\ket{\text{VAC}}$ is given by $\langle g(x) \rangle = {N\over 2L^{d}}\sum_{k,k'}\overline{c}_{k}c_{k'}(k+k')e^{-i(k-k')^{T}x}$. Therefore,
\begin{eqnarray}
\langle g(x) \rangle_{U[mv(x)_{j}]\ket{\psi_{k=0}}} &=& \text{Re} {iNm\over L^{2d}}\sum_{k}\int d^{d}x'\, v(x')_{j}x_{j}'\left( i\del_{x_{i}'}e^{-ik^{T}x'}\right)e^{ik^{T}x} + \mathcal{O}(v^{2}) \nonumber \\
&=& {Nm\over L^{d}}\left( x_{j}\del_{x_{i}}v(x)_{j} + v(x)_{j}\delta_{i,j} \right) + \mathcal{O}(v^{2})
\label{eqn:becproof}
\end{eqnarray}
where the second line follows from integration by parts. Taking the limit in Eq.(\ref{eqn:becproof}) as prescribed in Eq.(5) gives Eq.(6).

\section{Proof of Eq.(8) and relation to the winding number formula for $\rho_{s}$\label{sec:windingproof}}

From the unitary action in Eq.(1), an LGT in the $j$-th direction transforms a normal ordered Hamiltonian $H$ describing the dynamics of bosons of mass $m$ interacting pairwise with potential $V(x-x' )$ to the following Hamiltonian
\begin{eqnarray}
U[mv(x)_{j}]HU[mv(x)_{j}]^{\dagger}&\approx & {1\over 2m}\int d^{d}x\, \nabla  \psi(x)^{\dagger}\cdot \nabla  \psi(x) \nonumber \\
&-& \int d^{d}x\, g(x)^{T} \nabla (v(x)_{j}x_{j}) \nonumber \\
&+&{m\over 2} \int d^{d}x\, \Vert \nabla (v(x)_{j}x_{j} ) \Vert^{2}\psi(x)^{\dagger}\psi(x)\nonumber \\
&+& \int d^{d}xd^{d}x'\, V(x-x')\psi(x)^{\dagger }\psi(x')^{\dagger}\psi(x)\psi(x')
\label{eqn:lgtham}
\end{eqnarray}
where we have neglected terms in $\mathcal{O}(v^2)$, $\mathcal{O}(v\del v)$. It follows from this LGT that
\begin{equation}
-{\delta \Delta F \over \delta \del_{i}(v(x)_{j}x_{j})} \approx \langle  g(x)_{i} \rangle_{\rho(\beta)_{v(x)_{j}}} -m\del_{i}(v(x)_{j}x_{j})\langle \psi(x)^{\dagger}\psi(x) \rangle_{ \rho(\beta)_{v(x)_{j}} }.
\label{eqn:i1}
\end{equation}
Eq.(4) is now used to relate $\langle  g(x)_{i} \rangle_{\rho(\beta)_{v(x)_{j}}}$ to $\rho_{n}(x)_{i,j}$.  Taking the appropriate functional derivative of the equation $ \Delta F \approx  {m^{2}\over 2N!\beta Z} \sum_{s\in \mathfrak{S}_{N}}\mathbb{E}_{s}W_{s}(\lbrace x^{(\ell)} \rbrace)^{2}$ derived in the main text, and setting it equal to the right hand side of Eq.(\ref{eqn:i1}) results in Eq.(8).$\square$

Let us consider $\rho_{n}(x)_{j,j}$, which is relevant for isotropic systems. In this case, the local two fluid relation $\rho_{s}(x)_{j,j}+\rho_{n}(x)_{j,j}=m\langle \psi(x)^{\dagger}\psi(x) \rangle$ implies that Eq.(\ref{eqn:i1}) becomes \begin{equation}
-{\delta \Delta F \over \delta \del_{i}(v(x)_{j}x_{j})} \approx -\rho_{s}(x)_{j,j}v(x)_{j} -mx_{j} \del_{j}v(x)_{j}\langle \psi(x)^{\dagger}\psi(x) \rangle_{ \rho(\beta)_{v(x)_{j}} }.
\end{equation} The second term in the above equation vanishes if $v(x)_{j}=v_{j}=\text{const}$ for all $x$ and, therefore, in the special case of a constant velocity field, Eq.(8) implies that
\begin{equation}
\rho_{s}(x)_{j,j}\approx {m^{2}\over 2\beta N! Z \hbar^{2}v_{j}}{d \over dv_{j}} \sum_{s\in\mathfrak{S}_{N}}\mathbb{E}_{s}W_{s}(\lbrace x^{(\ell)} \rbrace)_{j}^{2}
\end{equation}
where, for $v(x)_{j}=v_{j}$, $W_{s}(\lbrace x^{(\ell)} \rbrace)_{j} = v_{j}\sum_{\ell=1}^{N}\left( x_{j}^{(\ell)}-x_{j}^{(s(\ell))}\right)$.
Taking $d=3$, and averaging ${1\over 3}\sum_{j=1}^{3}\rho_{s}(x)_{j,j}$ gives the well known result Eq.(22) of Ref.[39], independent of $x$.

\section{cMPS for states of the form Eq.(11)\label{sec:cmpsapp}}

We define the displacement operator for cMPS by $\mathcal{D}(Q,R)=\mathcal{P}\exp \left[ \int_{-{L\over 2}}^{{L\over 2}}dx\, Q(x)\otimes \mathbb{I} +  R(x)\otimes \psi(x)^{\dagger} \right]$, where $Q(x)$ and $R(x)$ are $D\times D$ matrix-valued distributions, $D$ is the \textit{bond dimension}, and $\mathcal{P}$ is the path-ordering operator.
For any $n\in \mathbb{Z}_{\ge 0}$, the state $a^{\dagger n}_{\ket{\psi}}\ket{\text{VAC}}$ can be written as 
$\del_{\xi}^{n} \text{Tr}\mathcal{D}(Q,\xi R) \ket{\text{VAC}} \big\vert_{\xi=0}$,
where $R(x)=\psi(x)$ is the wavefunction corresponding to $\ket{\psi}$ and $Q(x)=\lambda$ for any $\lambda \in \mathbb{R}$ (i.e., the bond dimension of the generating cMPS is $D=1$), and $\xi \in \mathbb{C}$. For the purposes of the present appendix, $Q(x)$ can be gauged away [43]. Analogously, given a collection $\lbrace \ket{\psi_{j}} \rbrace_{j=1,\ldots ,M}$, and defining $R_{j}(x)=\psi_{j}(x) \in L^{2}([-L/2,L/2])$, one has the following generating functional form of $\prod_{j=1}^{M}a^{\dagger n_{j}}_{\ket{\psi_{j}}}\ket{\text{VAC}}$ in terms of a non-translationally invariant cMPS, again with $D=1$:
\begin{eqnarray}
\prod_{j=1}^{M}a^{\dagger n_{j}}_{\ket{\psi_{j}}}\ket{\text{VAC}} &=& \prod_{j=1}^{M}\del_{\xi_{j}}^{n_{j}}\prod_{j=1}^{M}\text{Tr}\mathcal{D}(\xi_{j}R_{j}) \ket{\text{VAC}} \Big\vert_{\lbrace \xi_{j} \rbrace =0} \nonumber \\ 
&=& \prod_{j=1}^{M}\del_{\xi_{j}}^{n_{j}}\text{Tr}\mathcal{D}(\tilde{R}_{\lbrace \xi_{j} \rbrace})\ket{\text{VAC}} \Big\vert_{\lbrace \xi_{j} \rbrace =0} 
\end{eqnarray}
where $\tilde{R}_{\lbrace \xi_{j} \rbrace}(x) = \sum_{j=1}^{M}\xi_{j}\psi_{j}(x)$ and $\tilde{Q}(x)=\sum_{j=1}^{M} \lambda_{j}$, $\lambda_{j}\in \mathbb{R}$.
\end{widetext}

\bibliography{phasebib.bib}

\end{document}